\documentclass[fleqn,10pt]{article}
\usepackage{epsfig}
\newcommand{\be}{\begin{equation}}
\newcommand{\ee}{\end{equation}}

\newcommand{\eq}[1]{Eq.\,(\ref{#1})}

\newcommand{\pbar}{\bar p}

\newcommand{\alphabold}{\mbox{\small\boldmath $\alpha$}}

\newcommand{\delchisq}{\Delta \chi^2_i(x_i;\alphabold)}
\newcommand{\delchi}{\Delta \chi^2_i}

\newcommand{\delchimax}{{\delchi}_{\rm max}}
\newcommand{\x}{(\nu/m)}
\newcommand{\y}{(\nu_0/m)}
\begin{document}
\setcounter{secnumdepth}{4}
\renewcommand\thepage{\ }
%
%
\begin{titlepage} 
%
\newcommand\reportnumber{1106} 
\newcommand\mydate{\today} 
\newlength{\nulogo} 
\settowidth{\nulogo}{\small\sf{UW Report  MADPH-04-XXXX}}
\title{
\vspace{-.8in} 
\hfill\fbox{{\parbox{\nulogo}{\small\sf{
NUHEP Report  \reportnumber\\
UW Report MADPH-05-1447\\
          \mydate}}}}
\vspace{0.5in} \\
{
Analyticity as a Robust Constraint on the LHC Cross Section
}}

\author{
M.~M.~Block\\
{\small\em Department of Physics and Astronomy,} \vspace{-5pt} \\ 
{\small\em Northwestern University, Evanston, IL 60208}\\
\vspace{-5pt}
\  \\
F.~Halzen
\vspace{-5pt} \\ 
{\small\em Department of Physics,} 
\vspace{-5pt} \\ 
{\small\em University of
Wisconsin, Madison, WI 53706} \\
\vspace{-5pt}\\
%
\vspace{-5pt}\\
%
}    
\vspace{.5in}
\vfill
\date {}
\maketitle
\begin{abstract}

It is well known that high energy data alone do not discriminate between asymptotic $\ln s$ and $\ln^2s$ behavior of $pp$ and $\bar pp$ cross sections. By exploiting high quality low energy data, analyticity resolves this ambiguity in favor of cross sections that grow asymptotically as $\ln^2s$. We here show that two methods for incorporating the low energy data into the high energy fits give numerically identical results and yield essentially identical tightly constrained values for the LHC cross section. The agreement can be understood as a new analyticity constraint derived as an extension of a Finite Energy Sum Rule.

\end{abstract}
\end{titlepage} 
%
\pagenumbering{arabic}
\renewcommand{\thepage}{-- \arabic{page}\ --}  

High precision low energy data represent a powerful constraint on the high energy behavior of hadronic cross sections via duality\cite{igi,newfroissart}. The low energy data can be separated into two energy regimes, the resonance region and a region with energies in excess of a laboratory energy $\nu_0$ where the resonances average into a featureless cross section in the sense of duality. These data represent powerful constraints on asymptotic fits to {\it high energy} data. Igi and Ishida\cite{igi} realized these constraints using a Finite Energy Sum Rule (FESR) which numerically averages the resonances, while Block and Halzen\cite{newfroissart} simply required that the high energy amplitudes fit both the experimental cross sections and their derivatives at the transition energy $\nu_0$.  Both methods discriminate between a $\ln s$ and $\ln^2 s$ asymptotic behavior of the asymptotic cross section, conclusively favoring the latter. They appear to be more selective than conventional fitting techniques\cite{Cudell}.

In this note we will show that the constraints of Block and Halzen\cite{newfroissart} derive from analyticity\cite{analyticity}, as does the FESR(2) of Igi and Ishida\cite{igi}. The purpose of this note is to show that they are in fact equivalent, as confirmed by fitting the two apparently very different methods to a {\em common data set} of $pp$ and $\bar pp$ cross sections\cite{sieve}.

Following Block and Cahn\cite{bc}, we describe the high energy data in terms of real analytic amplitudes
\begin{equation}
f_+=i\frac{p}{4\pi}\left\{A+\beta[\ln (s/s_0) -i\pi/2]^2+cs^{\mu-1}e^{i\pi(1-\mu)/2}-i\frac{4\pi}{\nu}f_+(0)\right\},\label{evenamplitude_gp}
\end{equation}
for the  crossing-even real analytic amplitude\cite{BH,compton} and
\be
f_-=-\frac{p}{4\pi}Ds^{\alpha -1}e^{i\pi(1-\alpha)/2},
\ee
for the crossing-odd real analytic amplitude. If $\alpha < 1$, it parameterizes the Regge behavior of a crossing-odd amplitude which vanishes at high energies. $A$, $\alpha$, $\beta$, $c$, $D$, $s_0$ and $\mu$ are real constants. The variable $s$ is the square of the center of mass system (cms) energy, $\nu$ is the laboratory energy and $p$ is the laboratory momentum. The real constant $f_+(0)$ is the subtraction constant\cite{bc} required at $\nu=0$.  From the optical theorem we obtain the total cross section
\be
\sigma^\pm= A+\beta\left[\ln^2 s/s_0-\frac{\pi^2}{4}\right]+c\,\sin(\pi\mu/2)s^{\mu-1}\pm D\cos(\pi\alpha/2)s^{\alpha -1}  \label{sigmatot}
\ee
with  $\rho$, the ratio of the real to the imaginary part of the forward scattering amplitude,  given by
\be
\rho^\pm={1\over\sigma_{\rm tot}}\left\{\beta\,\pi\ln s/s_0-c\,\cos(\pi\mu/2)s^{\mu-1}+\frac{4\pi}{\nu} f_+(0)\pm D\sin(\pi\alpha/2)s^{\alpha -1}\right\},\label{rhogeneral}
\ee 
where the upper(lower) sign refer to $pp$($\bar p p$) scattering. 

In the high energy limit, $s\rightarrow2m\nu$ where $m$ is the proton mass, \eq{sigmatot} and \eq{rhogeneral} can be written as
\begin{eqnarray}
\sigma_0\left(\frac{\nu}{m}\right)&{\!\!\! =\!\!\! }&c_0+c_1\ln\left(\frac{\nu}{m}\right)+c_2\ln^2\left(\frac{\nu}{m}\right)+\beta_{\cal P'}\left(\frac{\nu}{m}\right)^{\mu -1}\label{sigma0},\\
\sigma^\pm\left(\frac{\nu}{m}\right)&{\!\!\! =\!\!\! }&\sigma_0\left(\frac{\nu}{m}\right)
\pm\  \delta\left({\nu\over m}\right)^{\alpha -1},\label{sigmapm}\\
\rho^\pm\left(\frac{\nu}{m}\right)&{\!\!\! =\!\!\! }&{1\over\sigma^\pm}\left\{\frac{\pi}{2}c_1+c_2\pi \ln\left(\frac{\nu}{m}\right)-\beta_{\cal P'}\cot\left({\pi\mu\over 2}\right)\left(\frac{\nu}{m}\right)^{\mu -1}+\frac{4\pi}{\nu}f_+(0)
\pm \delta\tan\left({\pi\alpha\over 2}\right)\left({\nu\over m}\right)^{\alpha -1} \right\}\!\!,\label{rhopm}
\end{eqnarray}
where we have introduced the even high energy cross section $\sigma_0\left(\frac{\nu}{m}\right)$, as well as the relations $A = c_0 + \frac{\pi^2}{4}c_2 - \frac{c_1 ^ 2}{ 4c_2}$, $ s_0 = 2m ^ 2 e^{-c_1 / (2c_2)}$, $\beta=c_2$, $c = \frac{(2m^2)^{1 - \mu} } {\sin(\pi\mu/ 2)}\beta_{\cal P'}$ and $D=\frac{(2m^2)^{1-\alpha}}{\cos(\pi\alpha/2)}\delta$. This transformation linearizes  \eq{sigmapm} in the real  coefficients $c_0, c_1, c_2, \beta_{\cal P'}$ and $\delta$, convenient for a $\chi^2$ fit to the experimental total cross sections and $\rho$-values.  Invoking Regge behavior, Igi and Ishida\cite{igi} fixed $\mu=0.5$, which is the value we adopt in order to directly compare our method to their FESR(2) constraint. 

At the transition energy $\nu_0$ where we will match our high energy fits to the low energy data,  we define
\begin{eqnarray}
\sigma_{\rm even}(\nu_0)&=&\frac{\sigma^{+}(\nu_0)+\sigma^-(\nu_0)}{2}\nonumber\\
&=&c_0+c_1\ln\y+c_2\ln^2\y+\beta_{\cal P'}\y^{\mu-1},\label{nu0}
\end{eqnarray}
where $\sigma^+$($\sigma^-$) are the total cross sections for $pp$($\bar p p$) scattering. Using the Block and Halzen value\cite{newfroissart} for $\sigma_{\rm even}(\nu_0)$, i.e., $\sigma_{\rm even}=48.58$ mb at $\nu_0=7.59$ GeV, we obtain the constraint
\begin{eqnarray}
c_0&=& \sigma_{\rm even}(\nu_0)-c_1\ln\y-c_2\ln^2\y-\beta_{\cal P'}\y^{\mu-1}\nonumber\\
&=&48.58 -2.091c_1-4.371c_2 -0.3516\beta_{\cal P'}\label{intercepteven}.
\end{eqnarray}
In brief, we have used the $\bar pp$ and $pp$ cross sections at the transition energy $\nu_0=7.59$ GeV to anchor the asymptotic fit to the low energy data. The precise choice of $\nu_0$ is not critical, as we will see further on.  We will actually show further on that \eq{intercepteven} is a consequence of analyticity.

To summarize, our strategy is to exploit the rich sample of low energy data just  above the resonance region, but well  below the energies where data are used in  our  high energy fit. At the transition energy $\nu_0$,  the experimental cross sections $\sigma_{\bar pp}(\nu_0)$ and $\sigma_{pp}(\nu_0)$ are used to determine $\sigma_{\rm even}(\nu_0)$ of \eq{intercepteven}. In turn, this constrains the asymptotic high energy fit so that it {\em exactly} matches the low energy data at the transition energy $\nu_0$, constraining the value of $c_0$  in \eq{intercepteven}. Local fits are made to data in the vicinity of $\nu_0$ in order to evaluate the cross sections that are introduced in the above constraint equation, \eq{intercepteven}. We next impose the constraint \eq{intercepteven} on a $\chi^2$ fit to Equations ({\ref{sigmapm}) and (\ref{rhopm}). For safety, we start the  data fitting at much higher  energy,  $\nu_{\rm min}=18.72$ GeV ($\sqrt {s_{\rm min}}=6$ GeV), well above $\nu_0$.
 
Given the previous analyses\cite{igi,newfroissart} we only consider an asymptotic $\ln^2s$ fit; the even amplitude parameter $c_0$ is  constrained  by \eq{intercepteven}, {\em i.e.,} by  $c_1$,  $c_2$ and $\beta_{\cal P'}$ and the experimental value of $\sigma_{\rm even}(\nu_0)$. We then perform a simultaneous fit to the experimental high energy values of $\sigma_{\bar pp},\sigma_{pp},\rho_{\bar pp}$ and $\rho_{pp}$ using six parameters: the even parameters $c_1$, $c_2$, $\beta_{\cal P'}$ and $f_+(0)$ and the odd parameters $\delta$ and $\alpha$. Only the first 3 parameters are needed to describe the cross section.

We now derive a  new constraint from analyticity\cite{analyticity}, closely following Igi and Ishida's\cite{igi}
 derivation  of the fixed-energy sum rule\footnote{We have changed their notation to conform with that used in the present paper, replacing  $F$ by $f$, the energy ${N}$ by $\nu_0$, and using capital letters for the real coefficients of their high energy parametrization , i.e.,  letting $c_0\rightarrow C_0,\ c_1\rightarrow C_1,\ c_2\rightarrow C_2$ and $\beta_{\cal P'}\rightarrow B_{\cal P'}$. In what follows, $m$ is the proton mass, $p$ is the laboratory momentum and $\nu$ is the laboratory energy.}
 FESR(2) which they used to constrain their  high energy fit. They wrote the imaginary part of their even high energy amplitude expressed in terms of their dimensionless coefficients $C_0,C_1,C_2,$ and $B_{\cal P'}$ as 
\be
{\rm Im}f_+(\nu)=\frac{\nu}{m^2}\left[C_0+C_1\ln\y+C_2\ln^2\y+B_{\cal P'}\y^{\mu-1}\right].
\ee
After using the optical theorem, it yields the same cross section as $\sigma_0$ of \eq{sigma0} in the high energy limit ($p\rightarrow \nu$), if we make the substitutions $c_0\rightarrow \frac{4\pi}{m^2}C_0,\ c_1\rightarrow \frac{4\pi}{m^2}C_1,\ c_2\rightarrow \frac{4\pi}{m^2}C_2,$ and $\beta_{\cal P'}\rightarrow \frac{4\pi}{m^2}B_{\cal P'}$; note that the Igi and Ishida coefficients are dimensionless, whereas our coefficients $c_0,\ c_1,\  c_2,$ and $\beta_{\cal P'}$ have dimensions of mb.  We next introduce the true even  amplitude  $F_+(\nu)$ (which, of course, we do not know),  along with the odd super-convergent difference amplitude\cite{dolen-horn-schmid} 
\begin{eqnarray}
\nu \tilde f_+(\nu)&\equiv&\nu\left[F_+(\nu)-f_+(\nu)\right]\label{nuftilde},
\end{eqnarray} 
which, because of  analyticity\cite{dolen-horn-schmid},   satisfies the odd super-convergence relation 
\be
\int_0^\infty\nu \,{\rm Im}\tilde f_+(\nu)\,d\nu=0.\label{super}
\ee
Note that $\nu\tilde f_+(\nu)$ satifies \eq{super}, although neither the odd amplitude $\nu F_+(\nu)$ nor the odd amplitude $\nu f_+(\nu)$ alone necessarily satisfies it.
Using \eq{super}}, we now write
\be
\int_0^\infty\nu \,{\rm Im}F_+(\nu)\,d\nu=\int_0^\infty\nu \,{\rm Im}f_+(\nu)\,d\nu.\label{super2}
\ee

Up until now, we have only used analyticity and the fact that the odd amplitude $\nu\tilde f_+(\nu)$ is super-convergent.  However, super-convergence of $\nu\tilde f_+(\nu)$ also implies that $f_+(\nu)$, if it is a {\em good} representation of the high energy behavior, approaches sufficiently close to $F_+(\nu)$ at some transition energy $\nu_0$---taken as an energy somewhat above the resonance region---that the difference between these amplitudes is neglectable. Hence, the integrand of \eq{super} is essentially  zero for energies above $\nu_0$, allowing us to truncate the integration of \eq{super2} at $\nu_0$, thus obtaining the finite energy relation  
\be
\int_0^{\nu_0}\nu \,{\rm Im}F_+(\nu)\,d\nu=\int_0^{\nu_0}f_+(\nu)\,d\nu=0\label{finite}
\ee
for sufficiently large $\nu_0$.

Applying the optical theorem to \eq{finite}, we have the relation
\begin{eqnarray}
\int_0^m\nu \,{\rm Im}F_+(\nu)\,d\nu + \frac{1}{4\pi}\int_m^{ {\nu_0}} \nu p\sigma_{even}(\nu)\,d\nu&=&\int_0^{\nu_0}\frac{\nu^2}{m^2}\left[C_0+C_1\ln\x+C_2\ln^2\x\right.\nonumber\\
&&\quad\qquad\left. +B_{\cal P'}\y^{\mu-1}\right]\,d\nu.\label{lefthandintegral}
\end{eqnarray}
Because $\int_m^{ {\nu_0}} \nu p\sigma_{even}(\nu)\,d\nu=\int_0^{p_0} p^2\sigma_{even}(p)\,dp$,
where $p_0=\sqrt{{\nu_0}^2-m^2}$, this is the FESR(2) derived by  Igi and  Ishida which follows from analyticity, much as dispersion relations do. 

Again using the optical theorem, we now rewrite \eq{lefthandintegral} in a more general form as
\be
\int_0^m\nu \,{\rm Im}F_+(\nu)\,d\nu + \frac{1}{4\pi}\int_m^{ {\nu_0}} \nu p\sigma_{even}(\nu)\,d\nu=\frac{1}{4\pi}\int_0^{\nu_0}\nu^2 \sigma_0(\nu)\, d\nu.\label{FESR2}
\ee
Another constraint can be derived from the observation that above relation is satisfied for any ${\nu_0}$ in the energy region above  the resonance region, where the  cross section is smooth. In particular,  \eq{FESR2} is valid at ${\nu_0}+\Delta {\nu_0}$, where $\Delta {\nu_0}$ is small, i.e, $0<\Delta\nu_0\ll \nu_0$. Subtracting \eq{FESR2} evaluated at ${\nu_0}$ from \eq{FESR2} evaluated at ${\nu_0}+\Delta {\nu_0}$ yields
\be
\int_{\nu_0}^{{\nu_0}+\Delta {\nu_0}} p\nu\sigma_{even}(\nu)\,d\nu =\int_{\nu_0}^{{\nu_0}+\Delta {\nu_0}}\nu^2  \sigma_0(\nu)\, d\nu,\label{sigconstraint}
\ee
where $\sigma_{even}(\nu)$ is the value of the cross section at laboratory energy $\nu$ and $\sigma_0(\nu)$ is the cross section at $\nu$  obtained from the asymptotic amplitude. In the limit of $\Delta {\nu_0}\rightarrow 0$, \eq{sigconstraint} yields the analyticity constraint,
\begin{eqnarray}
\sigma_{even}({\nu_0})&=&\frac{{\nu_0}}{p_0}\times \sigma_0({\nu_0})\approx \sigma_0({\nu_0}),\label{sigma0constraint}
\end{eqnarray}
a relation good to $\sim 0.04$\% for ${p_0}= 10$ GeV, the value chosen by Igi and Ishida. Thus, we see that analyticity, which is the underlying fabric of the FESR(2) derived by Igi and Ishida\cite{igi} (\eq{FESR2}), {\em also} requires that the high energy even cross section  $\sigma_0(\nu_0)$ of \eq{sigma0} must match the low energy  experimental cross section $\sigma_{even}(\nu_0)$, if the parametrization of the high energy forward scattering amplitude is a {\em good} parametrization. Further, this result is {\em independent} of the value of the non-physical integral $\int_0^m\nu \,{\rm Im}F_+(\nu)\,d\nu $ needed to evaluate the FESR(2) in \eq{FESR2}, even if it is very large.

Igi and Ishida\cite{igi} anchored their fit using \eq{FESR2},  by numerically integrating $p^2\sigma_{\rm even}=p^2[\sigma_{\bar pp}+\sigma_{pp}]/2$ over the low energy resonance region below ${p_0}=10$ GeV to obtain
\begin{eqnarray}
\frac{1}{4\pi}\int_0^{p_0}p^2\sigma_{\rm even}(p)\,dp
&=&3403\pm20 {\rm \ GeV}\label{fesrigi}.
\end{eqnarray}

Neglecting the error in \eq{fesrigi} and approximating the left-hand integral (the integral over the non-physical region) in \eq{lefthandintegral} as 3.2 GeV,  they obtained the constraint 
\be
C_0= 8.87 -2.04C_1-4.26C_2-0.367B_{\cal P'},\label{fesroriginal}
\ee
or, changing their dimensionless coefficients into our coefficients  which have units of mb---by multiplying their coefficients by $\frac{4\pi}{m^2}$---we rewrite the Igi and Ishida  constraint as
\be
c_0= 49.28 -2.04c_1-4.26c_2-0.367\beta_{\cal P'}.\label{fesr}
\ee
This is the constraint that will be used in an alternative fit to the high energy data.

Before presenting our results, we comment on the ``sieved'' data that we will use for fitting\cite{newfroissart}. It uses all of the data in the Particle Data Group\cite{pdg} archive for $\bar pp$ and $pp$ total cross sections and $\rho$-values with energies $\sqrt s\ge 6$ GeV. A robust $\ln^2s$ fit was obtained which minimizes  the  Lorentzian squared\cite{sieve}, {\em before} imposing the ``Sieve'' algorithm. The algorithm then proceeds iteratively to rid the data sample of the outliers, based on a maximum cut on the individual $\chi^2$ of the $i^{\rm th}$ point, defined as $\Delta\chi^2_i$. Details are shown in ref. \cite{newfroissart,sieve}. A value of $\chi^2$/d.f.=5.657 was obtained for 209 degrees of freedom using the unscreened data\cite{pdg}. This is to be compared to a value of $\chi^2$/d.f.=1.095 for 184 degrees of freedom, when using a $\delchimax=6$ cut in the ``Sieve'' algorithm\cite{sieve}. The ``Sieve'' algorithm eliminated 25 points with energies $\sqrt s\ge6$ GeV (5 $\sigma_{pp}$, 5 $\sigma_{\pbar p}$, 15 $\rho_{pp}$), while changing the total renormalized $\chi^2$ from 1182.3 to 201.4. The 25 points that were screened out had a $\chi^2$ contribution of 980.9, an average value of 39.2. For a Gaussian distribution, about 3 points with $\delchi>6$ are expected, with a total $\chi^2$ contribution of slightly more than 18, not 980.9. This demonstrated the efficiency of the ``Sieve'' algorithm\cite{sieve} in excluding outliers\cite{newfroissart,sieve}. The same data set with $\delchimax=6$ and $\sqrt s\ge6$ GeV is used in the present analysis.

Table \ref{table:ppfitnew} shows the results of a 6 parameter $\chi^2$ fit constrained by FESR(2) and, alternatively, by the analyticity constraint that matches $\sigma_{\rm even}$ at $\nu_0$. The resulting $\chi^2$ have been renormalized\cite{sieve} for the cut $\delchisq=6$. Both fits are excellent, each with a renormalized $\chi^2$ per degree of freedom slightly less than 1.
 
\begin{table}[h,t]                   
%
\def\arraystretch{1.15}            
\begin{center}				  
\begin{tabular}[b]{|l||c|c||}

\hline
\multicolumn{1}{|c||}{Parameters }
      &\multicolumn{2}{|c||}{$\sigma\sim \ln^2s$, $\delchimax=6$}\\ 
\cline{2-3}
	\multicolumn{1}{|c||}{}
      &\multicolumn{1}{c|}{FESR2 Fit}&\multicolumn{1}{c||}{``Analyticity''  Fit} \\
      \hline
	\multicolumn{3}{|c||}{\ \ \ \ \  Even Amplitude}\\
	\cline{1-3}
      $c_0$\ \ \   (mb)&$36.68$ &$36.95$\\ 
      $c_1$\ \ \   (mb)&$-1.293\pm0.151$ &$-1.350\pm0.152$\\ 
	$c_2$\ \ \ \   (mb)&$0.2751\pm0.0105$&$0.2782\pm0.105$\\
      $\beta_{\cal P'}$\ \   (mb)&$37.10$ &$37.17$\\ 
      $\mu$&$0.5$ &$0.5$\\ 
	$f(0)$ (mb GeV)&$-0.075\pm0.67$&$-0.073\pm 0.67$\\
      \hline
	\multicolumn{3}{|c||}{\ \ \ \ \  Odd Amplitude}\\
	\hline
      $\delta$\ \ \   (mb)&$-24.67\pm 0.97$ &$-24.42\pm 0.96$\\
      $\alpha$&$0.451\pm 0.0097$ &$0.453\pm 0.0097$\\ 
	\cline{1-3}
     	\hline
	\hline
	$\chi^2_{\rm min}$&158.2&157.4\\
	${\cal R}\times\chi^2_{\rm min}$&180.3&179.4\\ 
	degrees of freeedom (d.f).&181&181\\
\hline
	${\cal R}\times\chi^2_{\rm min}$/d.f.&0.996&0.992\\
\hline
\end{tabular}
     \caption{\protect\small The fitted results for a 6-parameter $\chi^2$ fit with $\sigma\sim\ln^2(s)$ and the cut $\delchimax=6$, for the FESR constraint $c_0= 49.28 -2.04c_1-4.26c_2-0.367\beta_{\cal P'}
$ and the ``analyticity'' constraint $c_0=48.58 -2.091c_1-4.371c_2 -0.3516\beta_{\cal P'}$. The renormalized\cite{sieve} $\chi^2_{\rm min}$/d.f.,  taking into account the effects of the $\delchimax$ cut, is given in the row  labeled ${\cal R}\times\chi^2_{\rm min}$/d.f. The errors in the fitted parameters have been multiplied by the appropriate $r_{\chi2}$ (see ref. \cite{sieve}).  \label{table:ppfitnew}}
\end{center}
\end{table}
\def\arraystretch{1}  

\begin{figure} 
\begin{center}
\mbox{\epsfig{file=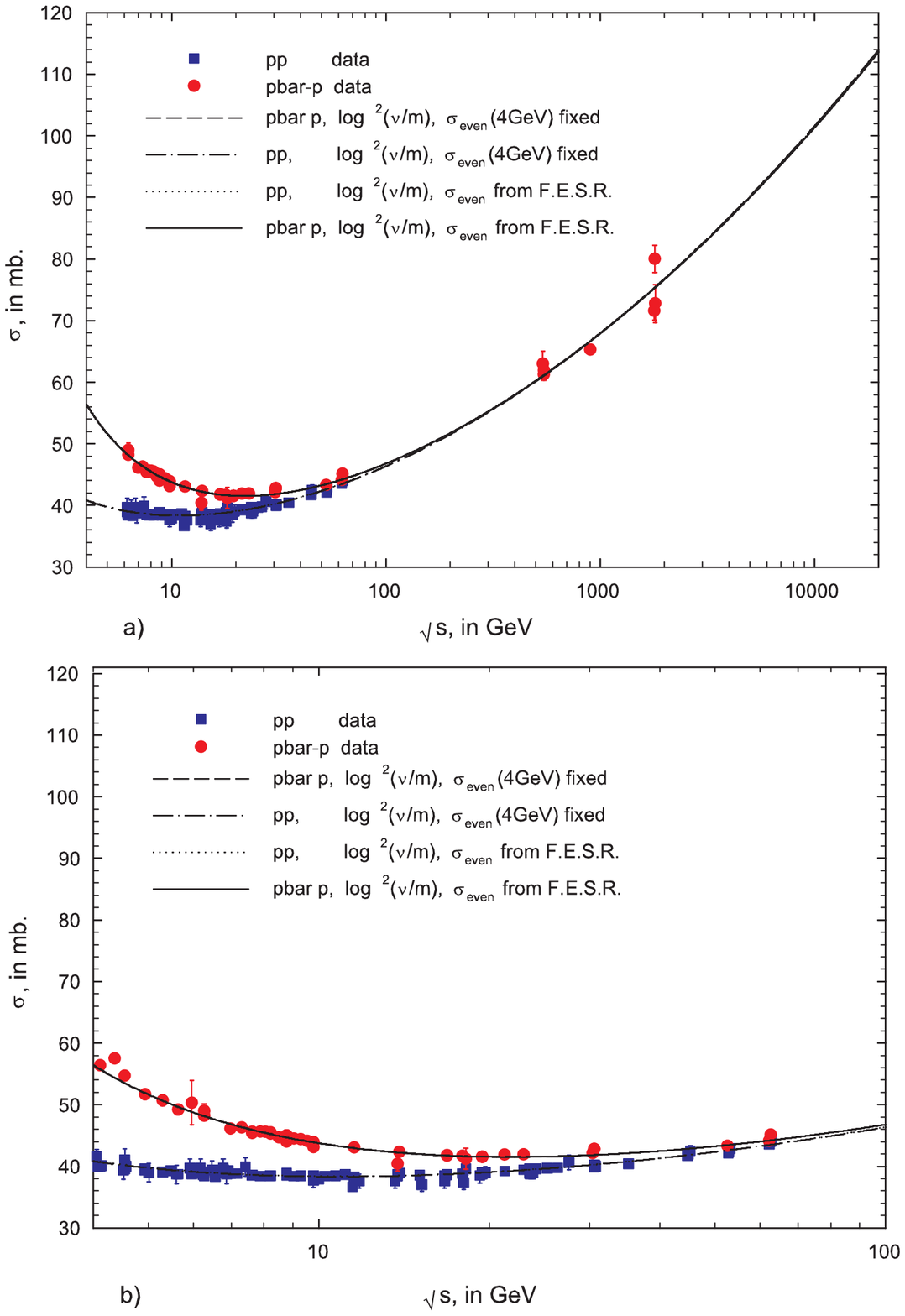,width=4.8in%
}}
\end{center}
\caption[]{ \footnotesize 
a) The fitted total cross sections $\sigma_{p p}$ and $\sigma_{\pbar p}$ in mb, {\em vs.} $\sqrt s$, in GeV, using the single constraint of Equations (\ref{intercepteven}) for the analyticity fit and (\ref{fesr}) for the FESR fit of Table \ref{table:ppfitnew}.  The circles are the sieved data  for $\pbar p$ scattering and the squares are the sieved data for $p p$ scattering for $\sqrt s\ge 6$ GeV.  The short dashed curve and dot-dashed curves are the analyticity fits---the {\em even} cross section at 4 GeV was fixed---to the $\bar p p$ and $pp$ data, respectively. The solid curve and dotted curves are the FESR fits to the $\bar p p$ and $pp$ data, respectively. It should be pointed out that the FESR and analyticity curves are essentially indistinguishable numerically for energies between 4 and 20000 GeV.\hspace{.1in} b) An expanded energy scale that additionally shows the cross section data that  exist\cite{pdg} between 4 GeV, where $\sigma_{\rm even}$ was fixed, and 6 GeV, the beginning of the fitted data. It should be emphasized that {\em none} of the data between 4 and 6 GeV were used in the fits. We note that  that the fits go through all of the unused points, with the exception of the $\bar pp$ point at 4.2 GeV, which would have been excluded by the ``Sieve'' algorithm\cite{sieve} because of its large $\delchi$, had it been used. 
  }
\label{fig:sigmapp}
\end{figure}

The $\bar pp$ and  $pp$ cross sections derived from the parameters of  Table \ref{table:ppfitnew} are shown in Fig.  \ref{fig:sigmapp}a) as a function of the cms energy, $\sqrt s$, for both methods. The $\bar p p$ (circles) and $pp$ (squares) data shown are the ``sieved" set.  The short dashed and dot-dashed curves are the analyticity constraint fits to the $\bar p p$ and $pp$ data, respectively. The solid curve and dotted curves are the some for the FESR fit. The difference between the two fits is negligible over the energy interval $4\le\sqrt s\le 20000$ GeV; they agree to an accuracy of about 2 parts in 1000.  It should be emphasized that the FESR fit uses the experimental resonance data below $\sqrt s=4$ GeV for evaluating the constraint of \eq{fesr}, whereas the analyticity constraint fit uses the even cross section {\em at} $\sqrt s=4$ GeV for the evaluation of its constraint, \eq{intercepteven}, {\em i.e.}, {\em the alternative fits do not share any data}. Both strongly support  $\ln^2 s$ fits that saturate the functional growth of the Froissart bound. 

In Fig.  \ref{fig:sigmapp}b) we show all of the $\bar pp$ and $pp$  cross section data\cite{pdg}  in the cms energy interval $4$ to $6$ GeV, none of which was used in our high energy fit.  Inspection of Fig.  \ref{fig:sigmapp}b) reveals that we could have imposed the analyticity constraint anywhere from 4 GeV to 6 GeV without modifying the result. Thus, our conclusions do not depend on the choice of $\nu_0$, the transition energy used in \eq{nu0}.

\begin{figure} 
\begin{center}
\mbox{\epsfig{file=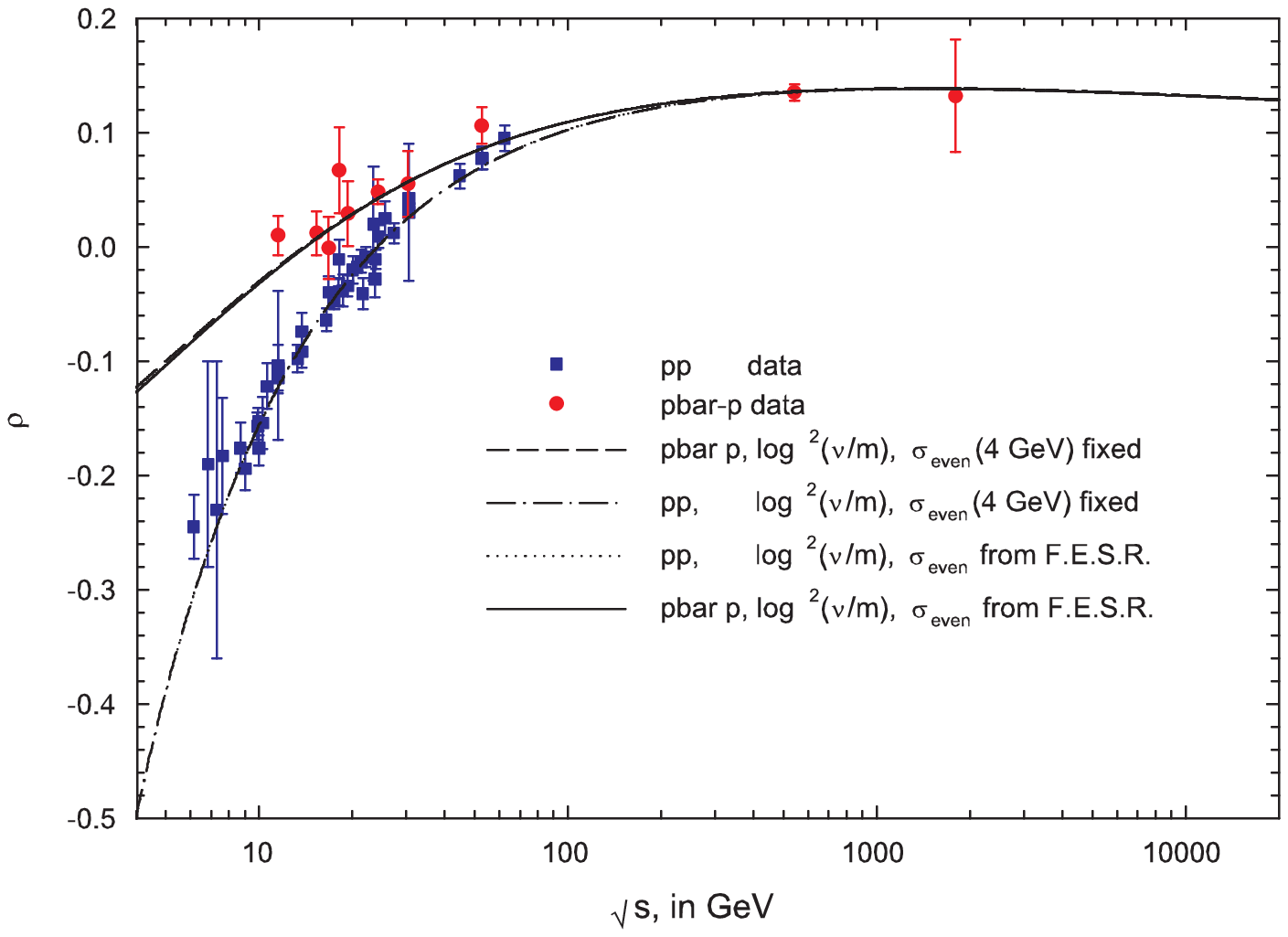,width=4.8in%
,bbllx=0pt,bblly=0pt,bburx=428pt,bbury=300pt,clip=%
}}
\end{center}
\caption[]{ \footnotesize 
 The fitted $\rho$-values, $\rho_{p p}$ and $\rho_{\pbar p}$, {\em vs.} $\sqrt s$, in GeV, using the single constraint of Equations (\ref{intercepteven}) for the analyticity fit and (\ref{fesr}) for the FESR fit of Table \ref{table:ppfitnew}.  The circles are the sieved data  for $\pbar p$ scattering and the squares are the sieved data for $p p$ scattering for $\sqrt s\ge 6$ GeV.  The short dashed curve and dot-dashed curves are the analyticity fits---the {\em even} cross section at 4 GeV was fixed---to the $\bar p p$ and $pp$ data, respectively. The solid curve and dotted curves are the FESR fits to the $\bar p p$ and $pp$ data, respectively. It should be pointed out that the FESR and analyticity curves are essentially indistinguishable numerically for energies between 4 and 20000 GeV.
  }
\label{fig:rhopp}
\end{figure}

Figure \ref{fig:rhopp} shows the fits for $\rho_{\bar pp}$ and $\rho_{pp}$ as a function of the cms energy $\sqrt s$; the ``sieved''  experimental data are shown for $\sqrt s \ge 6$ GeV. We conclude that the results are effectively the same for both fits and in good agreement with the experimental data. Accommodating $\rho$-values at lower energies allows one to constrain the cross section at higher energies by derivative dispersion relations, giving us additional confidence in our extrapolations.

Summarizing, the  FESR method and the new analyticity constraint introduced here yield fits to $\bar pp$ and $pp$ cross sections and $\rho$-values that agree to 2 parts in 1000 over the large energy interval $4{\rm \ GeV}\le\sqrt s\le 20000$ GeV. In particular, at the LHC energy of 14 TeV, the FESR fit predicts $\sigma_{pp}=107.2\pm 1.4$ GeV and $\rho_{pp}=0.130\pm0.002$, whereas the analyticity fit predicts $\sigma_{pp}=107.4\pm1.5$ GeV and $\rho_{pp}=0.131\pm0.002$. We showed that this agreement was expected---it is numerical confirmation that analyticity, in its two guises, gives identical numerical results. Further, the fact that the renormalized $\chi^2$ per degree of freedom in Table \ref{table:ppfitnew} is excellent, giving a high probability fit, means that the choice of our high energy even asymptotic amplitude  of \eq{sigma0} {\em satisfies} the analyticity constraint. It did not have to---had we used a poor representation for the even asymptotic amplitude, forcing the fit to go through the even cross section data at $\sqrt s=4$ GeV would have resulted in a very high $\chi^2$. This was demonstrated in references \cite{newfroissart} and \cite{igi}, where an asymptotic $\ln s$ parametrization was decisively rejected.

The fit of Block and Halzen\cite{newfroissart} which additionally constrains the cross section differences, as well as derivatives of the cross sections at $\sqrt s=4$ GeV, for both $pp$ and $p\bar p$, yields essentially the same cross section and $\rho$-value,  but with smaller errors.  Clearly, from analyticity considerations, this  technique is equivalent to evaluating additional FESRs, but is much more tractable numerically. This new tool yields   both robust and precise values for the total cross section at the LHC energy of 14 TeV, as well as at cosmic ray energies, allowing us to  make the prediction that at the LHC\cite{newfroissart}, $\rho_{pp}(14 {\rm \ TeV})=0.132\pm0.001$ and $\sigma_{pp}(14 {\rm \ TeV})=107.3\pm1.2$ mb.
 
\vspace{2mm}

\section*{Acknowledgments} The work of FH is supported  in part by the U.S.~Department of Energy under Grant No.~DE-FG02-95ER40896 and in part by the University of Wisconsin Research Committee with funds granted by the Wisconsin Alumni Research Foundation. One of us (MMB) would like to thank the Aspen Center of Physics for its hospitality during the writing of this paper.

\newpage

\end{document}